\documentclass{article}
\usepackage{geometry}

\usepackage{mathpple}
\usepackage{times}

\usepackage{amsmath}  
\usepackage{amssymb}  
\usepackage{mathrsfs} 

\usepackage{theorem}  
\usepackage{cite}     
\usepackage{comment}  

\usepackage{upref}
\usepackage{amsfonts}

\usepackage{verbatim}

\usepackage[dvipsnames,usenames]{color}
\usepackage{enumerate}

\usepackage{graphicx}

\usepackage{latexsym}

\usepackage{color}

\usepackage{multicol}
\usepackage{hhline}

\usepackage{psfrag}

\usepackage{tikz}
\usetikzlibrary{positioning}
\usepackage{pgfplots}

\usepackage{hhline}

\usepackage{algorithm}
\usepackage[noend]{algpseudocode}

\usepackage{xcolor,colortbl}

\usepackage{subcaption}

\newcommand{\be}[1]{\begin{equation}\label{#1}}
\newcommand{\ee}{\end{equation}}

\newcommand{\bc}{\begin{center}}
\newcommand{\ec}{\end{center}}

\newcommand{\cC}{{\cal C}}
\newcommand{\cD}{{\cal D}}

\newcommand{\bfa}{{\boldsymbol a}}
\newcommand{\bfb}{{\boldsymbol b}}
\newcommand{\bfc}{{\boldsymbol c}}

\newcommand{\bff}{{\boldsymbol f}}

\newcommand{\bfh}{{\boldsymbol h}}

\newcommand{\bfm}{{\boldsymbol m}}

\newcommand{\bfr}{{\boldsymbol r}}
\newcommand{\bfs}{{\boldsymbol s}}

\newcommand{\bfv}{{\boldsymbol v}}
\newcommand{\bfw}{{\boldsymbol w}}
\newcommand{\bfx}{{\boldsymbol x}}
\newcommand{\bfy}{{\boldsymbol y}}
\newcommand{\bfz}{{\boldsymbol z}}
\newcommand{\bfA}{{\mathbf A}}

\renewcommand{\le}{\leqslant}
\renewcommand{\leq}{\leqslant}

\renewcommand{\geq}{\geqslant}

\newcommand{\Cref}[1]{Co\-rol\-la\-ry\,\ref{#1}}

\theoremstyle{plain} \theorembodyfont{\normalfont\slshape}

\newtheorem{thm}{Theorem$\!$}
\newenvironment{theorem}{\begin{thm}\hspace*{-1ex}{\bf.}}{\end{thm}}

\newtheorem{prop}[thm]{Proposition$\!$}

\newtheorem{lem}[thm]{Lemma$\!$}
\newenvironment{lemma}{\begin{lem}\hspace*{-1ex}{\bf.}}{\end{lem}}

\newtheorem{cor}[thm]{Corollary$\!$}

\newtheorem{defi}[thm]{Definition$\!$}
\newenvironment{definition}{\begin{defi}\hspace*{-1ex}{\bf.}}{\end{defi}}

\newtheorem{claim}{Claim}
\newtheorem{cons}{Construction}

\theorembodyfont{\normalfont}

\newtheorem{remrk}{Remark$\!$}
\newenvironment{remark}{\begin{remrk}\hspace*{-1ex}{\bf .}}{\end{remrk}}

\definecolor{Codecolor}{named}{White}  

\newcommand{\Copen}{\mbox{\{\kern-5.50pt\{}}
\newcommand{\Cclose}{\mbox{\}\kern-5.50pt\}}}
\newcommand{\Cslash}{\mbox{$\backslash\kern-6.02pt\backslash$}}

\newtheorem{innercustomgeneric}{\customgenericname}
\providecommand{\customgenericname}{}
\newcommand{\newcustomtheorem}[2]{%
\newenvironment{#1}[1]
{%
\renewcommand\customgenericname{#2}%
\renewcommand\theinnercustomgeneric{##1}%
\innercustomgeneric
}
{\endinnercustomgeneric}
}
\newcustomtheorem{customproposition}{Proposition}
\newcustomtheorem{customtheorem}{Theorem}
\newcustomtheorem{customlemma}{Lemma}
\newcustomtheorem{customclaim}{Claim}
\newcustomtheorem{customcorollary}{Corollary}
\usepackage{pdfpages}

\begin{document}

\def\AllDefinitions{0}

\title{Near Optimal Code Construction for the Adversarial Torn Paper Channel With Edit Errors}

\date{}
\author{Maria Abu-Sini and Reinhard Heckel \\
School of Computation, Information and Technology, Technical University of Munich}
\maketitle

\thispagestyle{empty}

\begin{abstract}
Motivated by DNA storage systems and 3D fingerprinting, this work studies the adversarial torn paper channel with edit errors. This channel first applies at most $t_e$ edit errors (i.e., insertions, deletions, and substitutions) to the transmitted word and then breaks it into $t+1$ fragments at arbitrary positions. 
In this paper, we construct a near optimal error correcting code for this channel, which will be referred to as a $t$-breaks $t_e$-edit-errors resilient code. This code enables reconstructing the transmitted codeword from the $t+1$ noisy fragments. Moreover, we study list decoding of the torn paper channel by deriving bounds on the size of the list (of codewords) obtained from cutting a codeword of a $t$-breaks resilient code $t'$ times, where $t' > t$.
\end{abstract}

\section{Introduction}

In this work we study the \emph{$t$-breaks $t_e$-edit-errors channel}, which applies at most $t_e$ edit errors (i.e., insertions, deletions, and substitutions) to the transmitted word, then breaks it into $t+1$ fragments at arbitrary positions. We call an error correcting code that enables reconstructing the transmitted word from such $t+1$ noisy fragments a \emph{$t$-breaks $t_e$-edit-errors resilient} code.

The motivation for this study arises from DNA storage systems. As observed by Gimpel et al.~\cite{GrassChallenges} and Meiser et al.~\cite{ChemistryRepair}, the DNA strands tend to break to small pieces due to the DNA decay process. In addition, reading and writing the data in these systems introduce insertions, deletions, and substitutions. Hence, in these systems, we need to recover the strand from noisy fragments.

The error-free $t$-breaks channel (for which $t_e=0$) was introduced by Wang et al.~\cite{Raviv2024} along with a corresponding $t$-breaks resilient code. 

Here we consider the $t$-breaks channel with errors and the main contribution of this paper is constructing a $t$-breaks $t_e$-edit-errors resilient code. As explained in the sequel, this is achieved using the techniques of the papers~\cite{IrmakDocExch,DocExchange_Many,DocExchange_Hauepler,Orlitsky}, and~\cite{Raviv2024}.

The second contribution of the paper is studying the list decoding problem for the torn paper channel.
List decoding has been studied for several errors, including substitutions as in the papers~\cite{Elias,BassalygoList,GuruswamiList,Johnson1,Johnson2} and~\cite{Wozencraft}, and insertions and deletions in the papers~\cite{Hayashi} and~\cite{AntoniaListDecoding}, and is motivated for our setup as follows. 
A $t$-breaks resilient code enables retrieving the word from at most $t$ breaks, and since in DNA storage systems and 3D fingerprinting the strands and the objects might break to more pieces, one may ask what if $t'>t$ cuts happen to a codeword of a $t$-breaks resilient code. In this case, determining the specific transmitted codeword might be impossible, yet a list of all potential codewords may be obtained. Section~\ref{Sec:ListDecod} investigates this problem and establishes bounds on the list size.

The motivation for list decoding is to connect the adversarial and probabilistic torn-paper models. In the probabilistic model (Shomorony et al.\cite{ShomoronyEveryDistribution,ShomoronyFirstTornPaper}), each bit breaks independently with probability $p$, so we expect about $pn$ breaks in a sequence of length $n$. We can therefore encode using an adversarial $$(pn+c)$$-break (and $t_e$-edit) resilient code for a suitable constant $c$, which succeeds with high probability since the number of breaks is typically $\le pn+c+1$. When more breaks occur, list decoding can return all codewords consistent with the received fragments. Thus, list decoding quantifies when adversarial codes remain useful in the probabilistic setting (small lists) versus when one needs schemes designed specifically for that regime\cite{NestedHashing,NestedVT,EmbeddedVT}.

In this paper, we consider the adversarial setting where breaks can happen at arbitrary locations. According to Gimpel at al.~\cite{GrassChallenges}, the probabilistic setting of the torn paper channel captures the DNA breaking process in a better way.
Indeed, codes for this setting were devised in the papers~\cite{NestedHashing},~\cite{NestedVT}, and~\cite{EmbeddedVT}, however, further rigorous code-rate analysis is required therein. In this paper we aim to correct combinations of edit errors in the torn paper channel.
Meanwhile, we do that under the adversarial setting solely. However, we plan to extend this study in a future work for the probabilistic setting as well.
Furthermore, another adversarial setting of the torn paper channel, in which the fragments' lengths are bounded from below by some value, was studied by Bar-Lev et al. and Yehezkeally et al. in papers~\cite{AdversarialTornPaper} and~\cite{GeneralizedTornPaper}, respectively. However, as discussed in the work~\cite{Raviv2024} limiting the number of breakings (instead of the fragments' lengths) has a wider range of applications including 3D fingerprinting. Further similar channels, such as shotgun sequencing and chop and shuffle, were studied in the papers~\cite{Lit1,Lit2,Lit12,Lit3,Lit4, Lit5,Lit6,Lit7,Lit8,Lit9,Lit10,Lit15,Lit14} as well.

Lastly, all of the paper's analysis may be extended to a channel that applies at most $t$ breaks (instead of exactly $t$).
Before presenting the paper's contributions in Sections~\ref{Sec:CodeConstruction},~\ref{sec:Randominzed},~\ref{sec:Explicit}, and~\ref{Sec:ListDecod}, the following Subsection introduces the notations used throughout the paper. Notice that some proofs are omitted from the paper due to the lack of space.

\subsection{Definitions and Preliminaries} \label{Subsec:Definitions}

First as the paper focuses on the binary case, we denote by $\{0,1\}^n$ the set of all length-$n$ binary words. Then, $\left\{0,1 \right\}^* = \bigcup_{n \geq 0} \{0,1\}^n$. Given a word $\bfx \in \{0,1\}^{n}$ we enumerate its bits starting from $1$, hence $\bfx = x_1 x_2 \cdots x_n$. Notice that $\left| \bfx\right|$ denotes the length of the word, while $\left|S \right|$ designates the size of the set $S$.
A word $\bfy$ is said to be a \emph{substring} of $\bfx$ if for some index $1\leq i\leq n-\left|\bfy \right|+1$, $\bfy = x_{i}x_{i+1}\cdots x_{i+\left|\bfy\right|-1}$. Such $\bfy$ will be denoted by $\bfx_{\left[i,i+|\bfy|-1 \right]}$. Furthermore, $\bfx | \bfy$ represents the concatenation of $\bfx$ and $\bfy$.
Throughout the paper, the $\log$ is always taken to the base $2$.

An $\ell$-repeat-free word is a word in which every length-$\ell$ sequence occurs at most once as a substring of the word. Similarly, an $\ell$-repeat-free code is a set of $\ell$-repeat-free codewords~\cite{RFCodes, UniversalFramework}.

\section{$t$-Breaks $t_e$-Edit-Errors Resilient Codes} \label{Sec:CodeConstruction}

Theorems~\ref{theorem:RandomCode} and~\ref{theorem:ExplicitCode} are the main results in this paper.

\begin{theorem} \label{theorem:RandomCode}
For $t,t_e,n \in \mathbb{N}$ where $t+t_e =  o \left( \frac{n}{\log n\log \log n} \right) 
$, there exists a length-$n$ randomized $t$-breaks $t_e$-edit-errors resilient code of redundancy $\Theta \left( \left(t+t_e \right) \log n \log \log n \right)$.
\end{theorem}

Theorem~\ref{theorem:RandomCode} is proved in Section~\ref{sec:Randominzed} by constructing a randomized $t$-breaks $t_e$-edit-errors resilient code $\cC^{b,e}_{t,t_e}$ of redundancy $\Theta \left( \left( t+t_e\right) \log n \log \log n \right)$. The term \emph{randomized} here means that the code is constructed by some random process. In particular, $\cC^{b,e}_{t,t_e}$ is provided without an efficient encoder that gets arbitrary information word and encodes it into a codeword.
Note that Theorem~\ref{theorem:RandomCode} extends the following result from the paper~\cite{Raviv2024}. This result was also provided in Theorem $5$ and Corollary $1$ in the work~\cite{Raviv2024_Arxiv}.

\begin{theorem}~\cite{Raviv2024} \label{theorem:RavivCodeExist}
For $t \hspace{-0.5ex} = \hspace{-0.5ex} o \hspace{-0.5ex} \left( \hspace{-0.3ex} \frac{n}{\log n \log \log n} \hspace{-0.3ex} \right)\hspace{-0.3ex}$, there exists a length-$n$ randomized $t$-breaks resilient code of redundancy $\Theta \left( t\log n \log \log n \right)$.
\end{theorem}

Theorem~\ref{theorem:RavivCodeExist} is proved by Wang et al.~\cite{Raviv2024} by constructing a randomized $t$-breaks resilient code $\cC^b_{t}$ of redundancy $ \Theta\left( t \log n \log \log n \right)$. Notice that the redundancy of $\cC^{b,e}_{t,t_e}$ aligns with the redundancy of $\cC^b_t $. Our next result is given in Theorem~\ref{theorem:ExplicitCode} and proved in Section~\ref{sec:Explicit}. Specifically, Section~\ref{sec:Explicit} provides an encoder that receives an arbitrary information word and encodes it into a codeword of a $t$-breaks $t_e$-edit-errors resilient code $\cC^{b,e,enc}_{t,t_e}$. However, providing the efficient encoder here comes with the price of increasing the redundancy of $\cC^{b,e,enc}_{t,t_e}$ to $\Theta \left( \left(t+t_e \right) \log n \log \frac{n}{t+t_e} \right) $.

\begin{theorem} \label{theorem:ExplicitCode}
For $m,t,t_e\in \mathbb{N}$ where $t+t_e=  o \left( \frac{m}{\log m\log \log m} \right) 
$, there exists a polynomial run-time encoder that encodes every length-$m$ word into a codeword of a $t$-breaks $t_e$-edit-errors resilient code of length $ n=m+\Theta \left(  \left( t+t_e \right) \log m \log \frac{m}{t+t_e} \right)$. The codewords can be reconstructed from the fragments with run-time complexity
\begin{itemize}
    \item polynomial of $n$ if it is guaranteed that insertions and deletions did not occur in the fragments, yet
    \item $\Theta \left(t! n^4 \right)$ otherwise.
\end{itemize}
\end{theorem}

The challenge of designing an efficient encoder was addressed by Wang et al.~\cite{Raviv2024_Practical} as well, where Theorem~\ref{theorem:RavivCodeExplicit} was established. Notice that Theorem~\ref{theorem:ExplicitCode} extends Theorem~\ref{theorem:RavivCodeExplicit}. In particular, for the case of breakings solely, i.e., when $t_e = 0$, a shorter codeword is constructed in Theorem~\ref{theorem:RavivCodeExplicit}.

\begin{theorem}~\cite[Theorem 2]{Raviv2024_Practical} \label{theorem:RavivCodeExplicit}
For $t = o \left( \frac{m}{\log^4 m} \right)$, there exists a 
polynomial run-time encoder that 
encodes every length-$m$ word into a codeword of a $t$-breaks resilient code of length $m+\omega \left( t \log^2 m\right)$. The codewords can be reconstructed from the fragments in polynomial run-time complexity as well.
\end{theorem}

In Theorem~\ref{theorem:RedLowBound} we derive a lower bound on the redundancy of a $t$-breaks $t_e$-edit-errors resilient code. By comparing this lower bound with the redundancy of $\cC^{b,e}_{t,t_e}$, we conclude that $\cC^{b,e}_{t,t_e}$ is indeed near optimal when $t = O \left( n^{1-\epsilon} \right)$ for $\varepsilon > 0$.

\begin{theorem} \label{theorem:RedLowBound}
Let $\cC \subseteq \{0,1\}^n$ be a $t$-breaks $t_e$-edit-errors resilient code. Then,
$$ n- \log |\cC| \geq \Omega \left( \left(t+t_e\right) \log \frac{n}{t+t_e} \right). $$
\end{theorem}

Theorem~\ref{theorem:RedLowBound} follows the same proof strategy as Theorem 1 in the paper~\cite{Raviv2024}, which states that the redundancy of any length-$n$ $t$-breaks resilient code is bounded from below by $\Omega \left( t\log \frac{n}{t} \right)$. 

Lastly, we emphasize that to tackle both edit errors and breakings, we integrate in Sections~\ref{sec:Randominzed} and~\ref{sec:Explicit} ideas from the works~\cite{DocExchange_Many,DocExchange_Hauepler}, and~\cite{IrmakDocExch} into the construction of $\cC^b_t$ from the paper~\cite{Raviv2024}.

\section{Proof of Theorem~\ref{theorem:RandomCode}} \label{sec:Randominzed}

As the construction of $\cC^{b,e}_{t,t_e}$ is based on the code $\cC^b_t$ from the work~\cite{Raviv2024},
Subsection~\ref{subsec:Literature} first reviews the paper~\cite{Raviv2024}, yet with a slight modification to its definitions. Namely, we review the work~\cite{Raviv2024} using the terms and definitions of the papers~\cite{DocExchange_Many} and~\cite{DocExchange_Hauepler} to later simplify integrating techniques from these papers.
Then, Subsection~\ref{subsec:KeyIdea} highlights the key ideas behind the code $\cC^{b,e}_{t,t_e}$ and Subsection~\ref{subsec:Randominzed} provides the detailed construction.

\subsection{Background on the Code $\cC^b_{t}$ as Constructed in the paper~\cite{Raviv2024}} \label{subsec:Literature}

\begin{figure}
\centering
\includegraphics[scale=0.45]{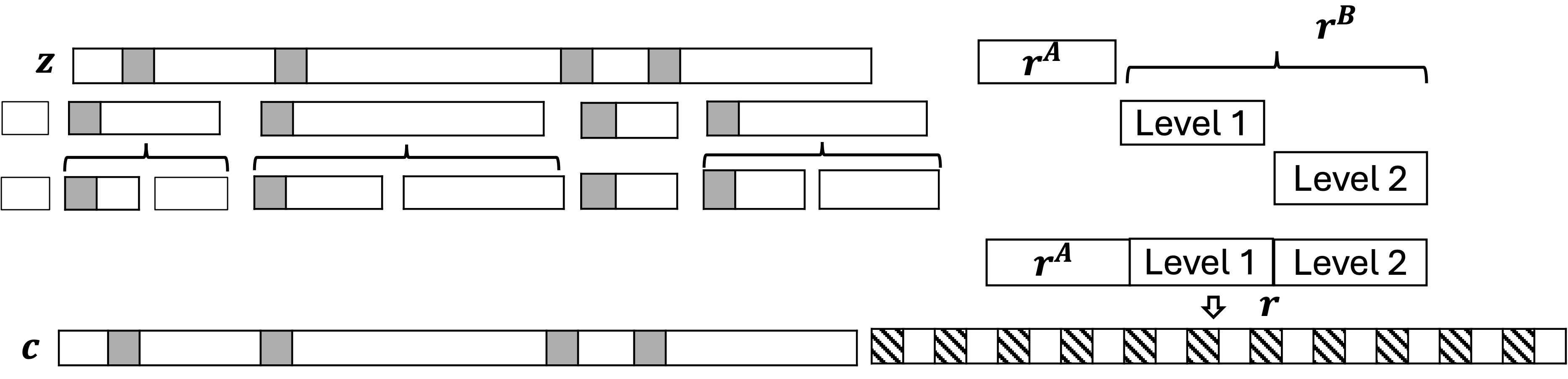}    
\caption{A codeword $\bfc \in \cC^b_{t}$ according to Construction~\ref{Cons:RavivCode} and Definitions~\ref{Def:LegitWord},~\ref{Def:GenerateRed}, and~\ref{Def:RedB}. $\bfz$ contains many markers. The leftmost part depicts the division to blocks in each level. $\bfr$ is obtained by inserting markers $M_{red}$ to $\bfr^A| \bfr^B$. Lastly, $\bfc = \bfz| \bfr$.}
\label{fig:RavivCodewordStructure}
\end{figure}

To prove Theorem~\ref{theorem:RavivCodeExist}, a randomized $t$-breaks resilient code $\cC^b_t$ is constructed in the papers~\cite{Raviv2024} and~\cite{Raviv2024_Arxiv}. This construction is summarized in the following and illustrated in Fig.~\ref{fig:RavivCodewordStructure}.

\begin{cons}\cite{Raviv2024} \label{Cons:RavivCode}
Let $\beta = 6$, $c \geq 3$ be some constant, and $m\in \mathbb{N}$. 
The following steps construct a $t$-breaks resilient code $\cC^b_t$ of length $m+\Theta \left( t \log m \log \log m\right)$.

\noindent\textbf{Step 1:} Let $\cC_{MU} \subseteq \{0,1\}^{ c 
\log m}$ be some mutually uncorrelated code of size $ \frac{m^c}{\beta c \log m} $ (which may be obtained from the work~\cite{Maya}). Choose some arbitrary $t+1$ words 
$M_{red} \triangleq \left\{ \bfm_0, \ldots, \bfm_{t+1} \right\} \subseteq \cC_{MU}.$ The words of $\cC_{MU}$ will be referred to as \emph{markers} due to the role they play as explained in the sequel.

\noindent\textbf{Step 2:} Let $\bfz \in \{0,1\}^m$ be a uniformly random word. If $\bfz$ is a \emph{legit word} (see definition~\ref{Def:LegitWord}), use the algorithm summarized in Definition~\ref{Def:GenerateRed} to generate redundancy $\bfr$. Otherwise, generate another word instead.

\noindent\textbf{Step 3:} The concatenation $
\bfz | \bfr$ constitutes a codeword.
\end{cons}

\begin{definition}~\cite{ Raviv2024} \label{Def:LegitWord}
Following the notations of Construction~\ref{Cons:RavivCode}, $\bfz \in \{0,1\}^m$ is said to be legit if it satisfies all of the following properties.
\begin{itemize}
    \item \textbf{Property $1$:} Every interval of length $2\beta c \log^2 m + c\log m -1$ contains a word of $\cC_{MU}$. 

    \item \textbf{Property $2$:} $\bfz$ is $\left( c\log m\right)$-repeat-free.

    \item \textbf{Property $3$:} $\bfz$ does not contain any marker from $M_{red}$.
\end{itemize}
\end{definition}

In Definitions~\ref{Def:GenerateRed} and~\ref{Def:RedB} we assume the markers occur in $\bfz$ at indices $p_1, \ldots, p_k$. Notice that the markers do not overlap in $\bfz$, for $\cC_{MU}$ is a mutually uncorrelated code.

\begin{definition} ~\cite{Raviv2024} \label{Def:GenerateRed}
The redundancy $\bfr$ is constructed as follows. \\
\textbf{Step 1:} 
First, define adjacency matrix $\bfA = \left[ A \right]_{\bfa, \bfb \in \cC_{MU}}$ of size $\left| \cC_{MU}\right| \times \left| \cC_{MU} \right|$ by
$$\bfA_{\bfa,\bfb} \hspace{-0.5ex} \triangleq  \hspace{-0.5ex} \begin{cases}
    p_{i+1} \hspace{-0.3ex} \hspace{-0.3ex}-p_{i}, & \hspace{-1ex}\text{if } \bfa, \bfb \text{ occurs at } p_i, p_{i+1}, \text{ respectively} \\
    0, &\hspace{-1ex} \text{otherwise}
\end{cases} \hspace{-0.5ex}. $$
$\bfA$ represents the distances between the markers in $\bfz$.
As explained by Wang et al.~\cite{Raviv2024}, each row in $\bfA$ may be represented using a length-$\left(2c\log m \right)$ binary word. Hence, by compA function from the paper~\cite{Raviv2024_Arxiv}, $\bfA$ may be represented using a vector $\bfv_{\bfA} 
\in \mathbb{F}_{2^{2c\log m}}^{ \left| \cC_{MU} \right| }$. Namely, $\bfv_{\bfA}$ is a length-$\left(  \left| \cC_{MU}\right| \cdot 2c\log m \right)$ binary vector, in which we consider sets of $2c\log m$ consecutive bits as elements in the extension field $\mathbb{F}_{2^{2c\log m}}$.  

\noindent\textbf{Step 2:} 
Encode $\bfv_{\bfA}$ using a systematic Reed-Solomon encoder to get $\Theta (t)$ redundant symbols in $\mathbb{F}_{2^{2c\log m}}$, these symbols are presented in a length-$\Theta \left(t \log m  \right)$ binary vector $\bfr^{A}$.

\noindent\textbf{Step 3:} Construct $\bfr^B$ as explained in Definition~\ref{Def:RedB}. 

\noindent\textbf{Step 4:} 
To obtain $\bfr$, insert markers from $M_{red}$ in $\bfr^A | \bfr^B$ in distances of $\left(c \log m \right) / 2$ bits in a specific pattern (which is omitted here due to the lack of space).
Since the markers are of length $c \log m$, then $\left| \bfr \right| \leq3 \left| 
\bfr^A | \bfr^B\right| /2$.
\end{definition}

\begin{definition}~\cite{Raviv2024_Arxiv,Raviv2024} \label{Def:RedB} Define \emph{level-$1$ blocks} to be $ \bfz_{[1,p_1-1]},$ $ \bfz_{[p_1, p_2-1]}, \ldots, \bfz_{[p_k,m]} $. In case $p_1=1$, then $\bfz_{\left[1, p_1-1 \right]}$ does not constitute a block.
Next, to generate level-$(\ell+1)$ blocks consider each level-$\ell$ block. If it is shorter than $c \log m$ then it will be a level-$(\ell+1)$ block as well. Otherwise, split it to halves, and these halves will be level-$(\ell+1)$ blocks instead.
The division stops when all blocks are of length less than $c \log m$.
For each block we assign some length-$(c\log m)$ binary hash value so that each two same-level blocks have distinct hash values. 
Denote the number of level-$\ell$ blocks by $N^{\ell}$, the level-$\ell$ blocks themselves by $\bfb^{\ell}_1, \bfb^{\ell}_2, \ldots, \bfb^{\ell}_{N^{\ell}} $, and their hash values by $\bfh^{\ell}_1, \bfh^{\ell}_2, \ldots, \bfh^{\ell}_{N^{\ell}}$.
By considering sets of consecutive $c\log m$ bits as elements in the extension field $\mathbb{F}_{2^{c\log m}}$, one can view $\bfh^{\ell}_1 | \cdots | \bfh^{\ell}_{N^{\ell}}$ as a vector in $\mathbb{F}_{2^{c\log m}}^{N^{\ell}}$, and denote it by $\bfh^{\ell} \triangleq \left( \bfh^{\ell}_1, \cdots , \bfh^{\ell}_{N^{\ell}}\right)\in \mathbb{F}_{2^{c\log m}}^{N^{\ell}} $. To construct $\bfr^B$, for every level $\ell\geq 2$ we encode $\bfh^{\ell}$ using systematic Reed-Solomon encoder to obtain $\Theta (t)$ redundant symbols in $\mathbb{F}_{2^{c\log m}}$, which can be represented in a length-$\Theta \left( t \log m\right)$ binary vector. These redundant bits are appended to $\bfr^B$. Overall, $\left| \bfr^B \right| = O \left( t \log m \log \log m\right)$, for level-$1$ blocks are of length $O\left( \log^2 m \right)$. Hence, there are $O \left( \log \log m\right)$ levels in $\bfz$.
\end{definition}

Lastly, the following lemma, which was also provided as Corollary $1$ in the paper~\cite{Raviv2024_Arxiv}, summarizes the redundancy of the code $\cC^{b}_{t}$.

\begin{lemma} \label{theorem:Red} ~\cite{Raviv2024}
The code $\cC^{b}_{t}$ of length $n$ has redundancy 
$O \left( t \log n \log \log n \right)$.
\end{lemma}

\subsection{Key Ideas Behind the Proofs of Theorems~\ref{theorem:RandomCode} and~\ref{theorem:ExplicitCode}} \label{subsec:KeyIdea}

To demonstrate the key ideas behind our proofs, we recall first how a codeword $\bfc \in \cC^b_t$ is reconstructed in the paper~\cite{Raviv2024} from its error-free fragments $\bff_1, \ldots, \bff_{t+1}$. To reconstruct $\bfc$ we iteratively go over all levels, and for each level $\ell$ we recover its hash values $\bfh^{\ell}_1, \ldots, \bfh^{\ell}_{N^{\ell}}$. Once a hash value $\bfh^{\ell}_i$ is known, we search for a fragment $\bff_j$ that has length-$\left| \bfb^{\ell}_{i} \right|$ substring with hash value $\bfh^{\ell}_i$. Then we affix $\bff_j$ to align its substring on the position of block $\bfb^{\ell}_i$. The hash values are chosen wisely so that every two substrings of $\bfc$ have distinct hash values, hence the affixing procedure recovers $\bfc$ correctly.

Next, assume the fragments $\bff_1,\ldots, \bff_{t+1}$ are obtained from at most $t_e$ edit errors. Then, the affixing procedure explained above is not guaranteed to succeed anymore, for an error in fragment $\bff_j$ might result in substring $\bfs_1$ with hash value $\bfh^{\ell}_1$, though $\bfb^{\ell}_1$ is not a block of $\bff_j$. Furthermore, two errors in $\bff_j$ might result in two consecutive non-overlapping substrings $\bfs_1$ and $\bfs_2$ that have the hash values $\bfh^{\ell}_1$ and $\bfh^{\ell}_{N^{\ell}}$, respectively. In this case, it is not clear where to affix $\bff_j$, at the beginning or at the end of $\bfc$. We resolve this, in Subsection~\ref{subsec:Randominzed}, using the idea of finding \emph{longest matching} from the papers~\cite{DocExchange_Many,DocExchange_Hauepler}, and~\cite{IrmakDocExch}. Indeed this technique gives a near optimal $t$-breaks $t_e$-edit-errors resilient code. However, it results in decoding run-time complexity $\Omega \left( (t+1)! \right)$ as it requires going over all possible fragments' concatenations.
Therefore, in the case of substitutions solely, as stated in Theorem~\ref{theorem:ExplicitCode}, we optimize the decoder and avoid going over all $(t+1)!$ concatenations. The optimized decoder is provided in the proof of Theorem~\ref{theorem:ExplicitCode} in Section~\ref{sec:Explicit}. In a future work, we plan to optimize the decoder in the presence of insertions and deletions too.

In Theorem~\ref{theorem:ExplicitCode} and Section~\ref{sec:Explicit}, to enable efficient encoding, 
we follow the strategy of Cheng et al.~\cite{DocExchange_Many} and Haeupler~\cite{DocExchange_Hauepler}
and 
reduce the number of level-$1$ blocks, which then increases the length of these blocks, and hence the number of all levels, and the overall redundancy as well.

\subsection{Construction of $\cC^{b,e}_{t,t_e}$} \label{subsec:Randominzed}

Construction~\ref{Cons:BreaksErrorsCode} modifies Construction~\ref{Cons:RavivCode} to obtain a $t$-breaks $t_e$-edit-errors resilient code $\cC^{b,e}_{t,t_e}$.

\begin{cons} \label{Cons:BreaksErrorsCode}
    Let $\beta =6, c\geq 5$ be some constant, and $m\in \mathbb{N}.$ The following steps construct a $t$-breaks $t_e$-edit-errors resilient code $\cC^{b,e}_{t,t_e}$ of length $m+\Theta \left( \left( t+t_e\right) \log m \log \log m \right)$.

    \noindent \textbf{Step 1:} Choose $\cC_{MU}$ and $M_{red}$ as in Construction~\ref{Cons:RavivCode}, yet now $M_{red} = \left\{\bfm_0, \ldots, \bfm_{36 \left(t+t_e \right)\log \log m} \right\}$ is of size $36 \left( t+t_e\right)\log \log m $.

    \noindent \textbf{Step 2:} Let $\bfz \in \{0,1\}^m$ be a uniformly random word, if it is not legit reselect it until a legit word is obtained. Let $\bfz' = \bfm_0 | \bfz$. Henceforth, we consider $\bfz'$ and assume the markers occur in it at indices $p_1, \ldots, p_k$, where $p_1 = 1$.

    \noindent \textbf{Step 3:} To obtain $\bfr^A$, encode $\bfv_{\bfA}$ (where $\bfA$ is the adjacency matrix of $\bfz'$) using a systematic Reed-Solomon encoder as in Definition~\ref{Def:GenerateRed}, yet now produce $4t+6t_e$ redundant symbols over $\mathbb{F}_{2^{2c\log m}}$. Hence, now $\left|\bfr^A\right| = \left( 8t+12t_e\right)c\log m$.

    \noindent \textbf{Step 4:} To generate $\bfr^B$, divide $\bfz'$ to blocks, where level-$1$ blocks are $\bfz'_{\left[p_1, p_2-1\right]}, \bfz'_{\left[p_2, p_3-1\right]}, \ldots, \bfz'_{\left[p_k,m\right]}$. Then, to generate level-$(\ell+1)$ block, level-$\ell$ block is divided only if it is longer than $3c\log m$. The hash value at the first level is the length-$(c\log m)$ prefix of the block (i.e., the marker). For other levels $\ell \geq 2$, $\bfh^{\ell}_i$ is length-$(3c\log m)$ prefix of the block if $\left| \bfb_i^{\ell} \right| \geq 3c\log m$, and $\bfb_i^{\ell} | 1 | 0^{3c\log m-1 - \left| \bfb^{\ell}_i \right|}$ otherwise.
    Now $\bfh^{\ell} = \left(\bfh^{\ell}_1, \ldots, \bfh^{\ell}_{N^{\ell}}\right) \in \mathbb{F}_{2^{3c\log m}}^{N^{\ell}}$. The redundancy for each level is obtained by encoding $\bfh^{\ell}$ for every $\ell \geq 2$ to get $2t+6t_e$ redundant symbols over $\mathbb{F}_{2^{3c\log m}}$. Thus, overall, $\left| \bfr^B \right| \leq \left( 2t+6t_e \right) \cdot 3c\log m \log \log m$.

    \noindent \textbf{Step 5:} Before inserting markers $M_{red}$,
    consider $\frac{c \log m}{2}$ consecutive bits in $\bfr^A | \bfr^B$ as symbols over the field $\mathbb{F}_{2^{ \left(c\log m\right) /2 }}$. Then, $\bfr^A | \bfr^B \in \mathbb{F}_{ 2^{\left(c\log m\right) /2 }}^{l}$, where $l \leq 30 \left(t+t_e \right) \log\log m$. Encode $\bfr^A | \bfr^B $ using a systematic Reed-Solomon code to obtain $t+6t_e$ redundant symbols over $ \mathbb{F}_{ 2^{\left(c\log m\right) /2 }}$, and represent these symbols using a length-$\frac { \left( t+6t_e \right)c \log m}{2}$ binary word $\bfr'$. Lastly, $\bfr$ is obtained by inserting markers from $M_{red}\setminus \left\{ \bfm_0 \right\}$ in a predetermined order in $\bfr^A | \bfr^B | \bfr'$ so that the markers are of distances $\left(c \log m \right)/2$.

    \noindent \textbf{Step 6:} The obtained codeword is $\bfc \hspace{-0.5ex} = \hspace{-0.5ex} \bfz' \hspace{-0.4ex} |  \hspace{-0.4ex}\bfr \hspace{-0.5ex} = \hspace{-0.5ex} \bfm_0 \hspace{-0.4ex} | \hspace{-0.4ex} \bfz \hspace{-0.4ex} | \hspace{-0.4ex} \bfr$.
\end{cons}

\begin{figure}
\centering
\includegraphics[scale=0.5]{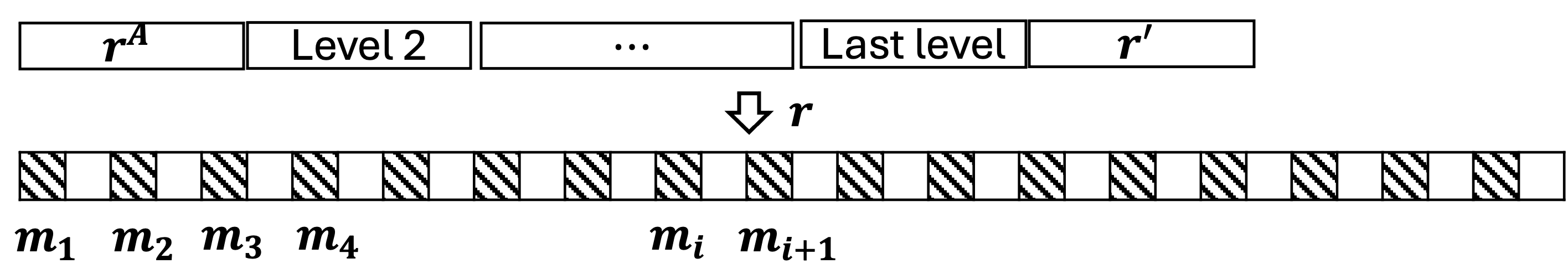}    \caption{Redundancy $\bfr$ in $\cC^{b,e}_{t,t_e}$.}
\label{fig:NewRed}
\end{figure}

Lemmas~\ref{lem:Red} and~\ref{lem:Decoding} establish Theorem~\ref{theorem:RandomCode}, where Lemma~\ref{lem:Red} may be proved by the same analysis as Corollary 1 in the paper~\cite{Raviv2024_Arxiv}.

\begin{lemma} \label{lem:Red}
Assume $\cC^{b,e}_{t,t_e}$ is of length $n$. Then it has redundancy
$O \left( \left(t+t_e \right) \log n \log \log n \right).$
\end{lemma}

\begin{lemma} \label{lem:Decoding}
It is possible to construct a codeword $\bfc \in \cC^{b,e}_{t,t_e}$ from any $t+1$ fragments obtained by applying at most $t_e$ edit errors. 
\end{lemma}

Remarks~\ref{Remark:EditErrorMarker} and~\ref{Remark:RepeatFree} are important for the proof of Lemma~\ref{lem:Decoding}.

\begin{remark} \label{Remark:EditErrorMarker}
Let $\bfc'$ be a word obtained from applying edit errors to $\bfc$. A set of errors that caused an occurrence of a marker in $\bfc'$ can not cause any other occurrence of a marker, for $\cC_{MU}$ is a mutually uncorrelated code. For instance, assume $\bfc = c_1 c_2 c_3 c_4 c_5 c_6 c_7$ and $\bfc' = 1 | c_1 \Bar{c}_2 c_3 c_5 c_6 c_7$, where $\Bar{c}_2$ denotes the complement of the bit $c_2$. If $1 | c_1 \Bar{c}_2 c_3 c_5 \in \cC_{MU}$, then the insertion of $1$, the substitution $c_2$, and the deletion of $c_4$ caused a new occurrence of a marker. They can not cause any other occurrence of a marker for $\cC_{MU}$ is a mutually uncorrelated code.
\end{remark}

\begin{remark} \label{Remark:RepeatFree}
Though $\cC^{b,e}_{t,t_e}$ is generated from legit words, which are $\left(c\log m\right)$-repeat-free, $\cC^{b,e}_{t,t_e}$ is not necessarily $\left(c\log m\right)$-repeat-free, yet it is $\left(3c\log m\right)$-repeat-free.
\end{remark}

\begin{figure}
\centering
\includegraphics[scale=0.5]{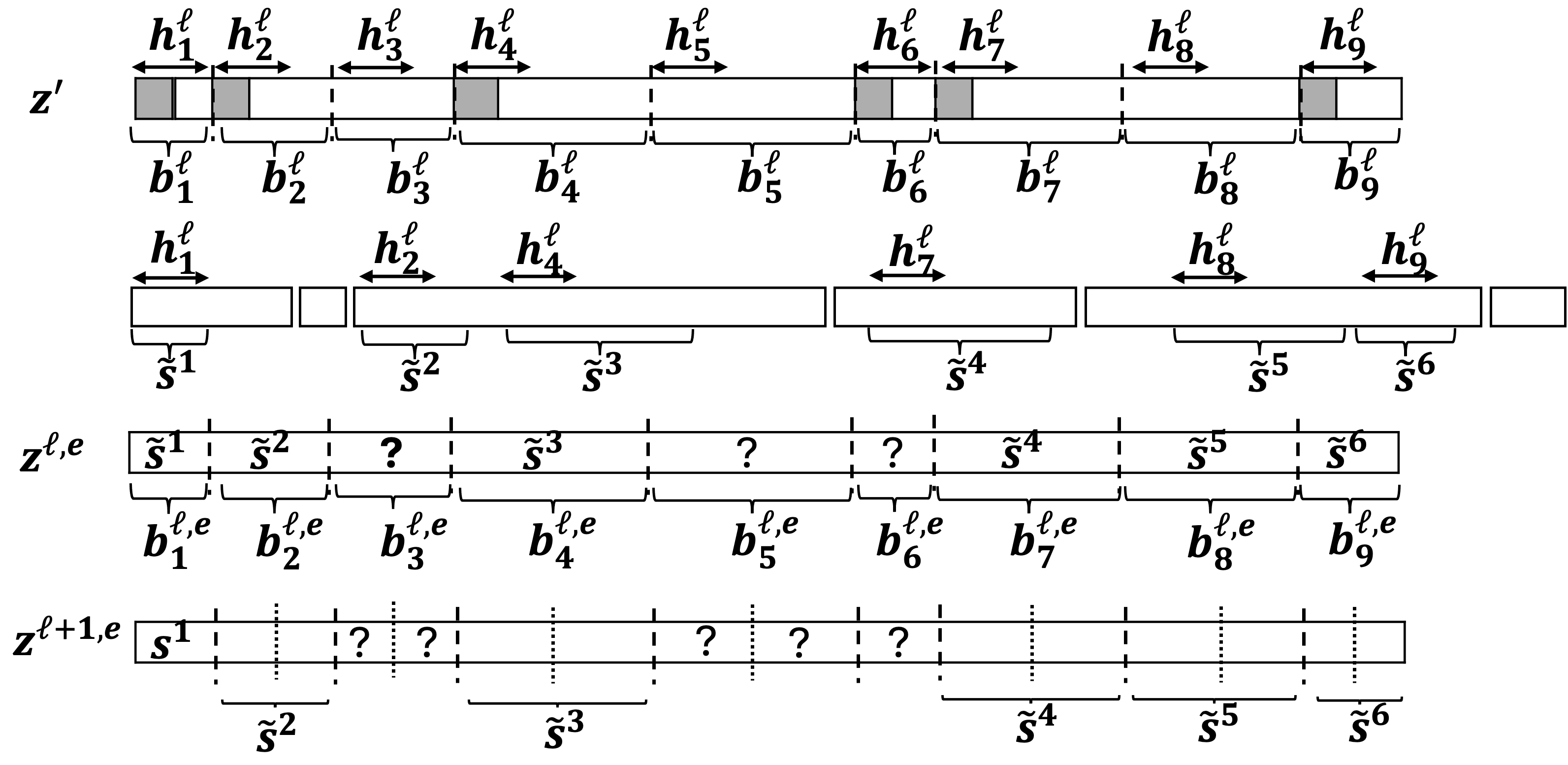}    

\caption{Matching between $\bfc$ and $\bfc^e$, and estimations $\bfz^{\ell,e}$ and $\bfz^{\ell+1,e}$, specifically in this case $N^{\ell}=9, \Tilde{k}=6, {\Tilde{j}}_1 =1, {\Tilde{j}}_2 = 2, {\Tilde{j}}_3 = 4, {\Tilde{j}}_4 = 7, {\Tilde{j}}_5 = 8$, and ${\Tilde{j}}_6 = 9$.}
\label{fig:Matching}

\end{figure}

\begin{proof} [Sketch proof of Lemma~\ref{lem:Decoding}]
    The following steps recover a codeword $\bfc = \bfz | \bfr \in \cC^{b,e}_{t,t_e}$ from noisy fragments $\bff_1, \ldots, \bff_{t+1}$ obtained by at most $t_e$ edit errors.

Step 1 -- Reconstructing $\bfr$: As depicted in Fig.~\ref{fig:NewRed},
$\bfr = \bfm_1 | \bfx_1  |  \cdots | \bfm_{l} |\bfx_{l},  $
where 
$\bfr^A | \bfr^B | \bfr' = \bfx_1 | \cdots | \bfx_{l }$ and $l \leq 36 \left( t+t_e \right)\log \log m$. 
To retrieve $\bfr$, Construct an estimation 
$ \bfx^e =  \bfx^e_1 | \cdots | \bfx^e_{l} $ by 
assigning $\bfx^e_i = ?$ to denote an erasure unless $\bfm_i$ occurs exactly once in all fragments, specifically in some fragment $\bff$ at index $h$, and $\bff$ is long enough to satisfy $h+ \left| \bfm_i \right| + \left| \bfx_i \right|-1 \leq \left| \bff \right|$.
In this case, assign $\bfx^e_i = {f}_{ h+\left| \bfm_i \right| } \cdots {f}_{h+\left| \bfm_i \right|+ \left| \bfx_i \right|-1}$.
Consider $\bfr^A | \bfr^B | \bfr'$ and $\bfx^e$ as vectors in $\mathbb{F}_{2^{\left( c\log m\right)/2}}^{l}$. Then,
symbol $\bfx^e_i$ is said to be erroneous if $\bfx^e_i \neq \bfx_i$.
By Remark~\ref{Remark:EditErrorMarker} there are at most $2t_e$, $t+2t_e$ errors, erasures in $\bfx^e$, respectively.
Therefore, applying the Reed Solomon decoder on $\bfx^e$ recovers $\bfr^A | \bfr^B | \bfr'$.

Step 2 -- Recovering the markers in $\bfz'$ and the indices $p_1, \ldots, p_k$: First, Step $1$ retrieves $\bfr^A$. Second, we construct an estimation $\bfA^e$ of the adjacency matrix $\bfA$ just as in the paper~\cite{Raviv2024}, i.e., for markers $\bfa, \bfb \in \cC_{MU} \setminus M_{red} $,
$$ \bfA^e_{\bfa, \bfb} = \begin{cases} 
d , & \hspace{-1.7ex} \begin{tabular}{l}
 if $\bfa$ occurs only in a  single fragment 
 $\bff$ and \\
 $\bfb$ occurs $d$ bits after it in the same 
 fragment
\end{tabular} \\
0 , & \hspace{-1.5ex} \hspace{1.2ex}\text{otherwise}
\end{cases} \hspace{-1.7ex}.$$
By Remark~\ref{Remark:EditErrorMarker} and the analysis of Wang et al.~\cite{Raviv2024}, 
$\bfA^e$ differs from $\bfA$ in at most $2t+3t_e$ rows.
Thus, the vectors $\bfv_{\bfA^e}, \bfv_{\bfA} \in \mathbb{F}_{2^{2c\log m}}^{\left|\cC_{MU} \right|}$ representing $\bfA, \bfA^e$, respectively, differ in at most $2t+3t_e$ symbols, and hence applying the Reed Solomon decoder on $\bfv_{\bfA^e} | \bfr^A$ recovers $\bfv_{\bfA}$.

Step 3 -- Recovering $\bfz$:  Following the papers~\cite{DocExchange_Many} and~\cite{DocExchange_Hauepler}, we will iteratively recover the hash values of each level. At the end, this gives $\bfz$, for the hash values in the last level are the complete blocks concatenated with a $1$ and a run of $0$'s. The hash values of the first level are the markers, hence are recovered in Step 2. 
Next, assume $\bfh^{\ell}_1, \ldots, \bfh^{\ell}_{N^{\ell}}$ are known. We find $\bfh^{\ell+1}_1, \ldots, \bfh^{\ell+1}_{N^{\ell+1}}$ by first considering every possible concatenation of the fragments
$\bfc^e = \bff_{i_1} | \cdots | \bff_{i_{t+1}}.$ 
For each (concatenation) $\bfc^e$
we find a longest possible sequence of non-overlapping substrings $\bfs^1, \ldots, \bfs^k$ in $\bfc^e$ that satisfies the following requirements.

\noindent\textit{Requirement 1: Each substring is wholly contained in some fragment.}

\noindent\textit{Requirement 2: For some indices $1 \leq j_1 < j_2 < \cdots < j_k \leq N^{\ell}$, the substrings $\bfs^1, \ldots, \bfs^k$ are of length $ \left| \bfb^{\ell}_{j_1}\right|, \ldots, \left| \bfb^{\ell}_{j_k} \right|$ and have hash values $\bfh^{\ell}_{j_1}, \ldots, \bfh^{\ell}_{j_k} $, respectively.}

Following Cheng et al.~\cite{DocExchange_Many} and Haeupler et al.~\cite{DocExchange_Hauepler}, the sequence $\bfs^1, \ldots, \bfs^k$ will be referred to as a \emph{matching} to indicate that we found matching blocks between $\bfc$ and $\bfc^e$ in terms of hash values.

Next denote a longest matching among all concatenations by $ {\Tilde{\bfs}}^1, \ldots, {\Tilde{\bfs}}^{\Tilde{k}} $, and denote its indices guaranteed from Requirement $2$ by $ \Tilde{j}_1, \ldots, \Tilde{j}_{\Tilde{k}} $.
Recall that 
$ \bfz' = \bfm_0 | \bfz = \bfb^{\ell}_1 | \cdots | \bfb^{\ell}_{N^{\ell}}$. As illustrated in Fig.~\ref{fig:Matching}, we construct an estimation of $\bfz'$, 
$\bfz^{\ell,e} = \bfb_1^{\ell,e} | \cdots | \bfb_{N^{\ell}}^{\ell,e} $, where for every $1\leq i \leq N^{\ell}$, $\left| \bfb_i^{\ell} \right| = \left| \bfb_i^{\ell,e} \right|$ and

$$ \bfb_i^{\ell,e} = \begin{cases}
{\Tilde{\bfs}}^{i'}, & \text{if } {\Tilde{\bfs}}^{i'} \text{ has hash value } \bfh_i^{\ell} \\
?, & \text{otherwise}
\end{cases}.
$$
Recall also that $\bfz' = \bfb^{\ell+1}_1 | \cdots | \bfb^{\ell+1}_{N^{\ell+1}}$. 
Hence, as depicted in Fig.~\ref{fig:Matching}, we construct an estimation $\bfz^{\ell+1,e} = \bfb_1^{\ell+1,e}| \cdots | \bfb_{N^{\ell+1}}^{\ell+1,e}$ of level-$\left(\ell+1 \right)$ blocks by dividing the blocks of $\bfz^{\ell,e}$ in the same way we divide the level-$\ell$ blocks in $\bfz'$. Here a block of erasure in $\bfz^{\ell,e}$ translates to at most two blocks of erasure in $\bfz^{\ell+1,e}$. Similarly, if $\bfb_h^{\ell,e} = \bfs$ for some $\bfs \in \left\{ {\Tilde{\bfs}}^1, \ldots, {\Tilde{\bfs}}^{\Tilde{k}} \right\}$, then the two halves of $\bfb_h^{\ell,e}$ get the prefix and suffix of $\bfs$. 
Lastly, construct estimation
$ \bfh^{\ell+1,e} = \left(\bfh_1^{\ell+1,e}, \ldots, \bfh_{N^{\ell+1}}^{\ell+1,e} \right) $
of $\bfh^{\ell+1} = \left(\bfh_1^{\ell+1}, \ldots, \bfh_{N^{\ell+1}}^{\ell+1} \right)$ by taking $\bfh^{\ell+1,e}_h$ to be the hash value of the block $\bfb^{\ell+1,e}_h$ in $\bfz^{\ell+1,e}$.
The following statements hold.
\begin{itemize}
\item By Claim~\ref{claim:Erasures}, there are at most $t+t_e$ erased blocks in $\bfz^{\ell,e}$, thus at most $2\left(t+t_e \right)$ erased blocks in $\bfz^{\ell+1,e}$.
\item By Claim~\ref{claim:Errors}, there are at most $t_e $ erroneous blocks in $\bfz^{\ell,e}$ (for which $\bfb^{\ell,e}_h \neq \bfb^{\ell}_h$), thus at most $2t_e$ erroneous blocks in $\bfz^{\ell+1,e}$ (for which $\bfb^{\ell+1,e}_h \neq \bfb^{\ell+1}_h$).
\end{itemize}
Recall that $\bfh^{\ell+1}, \bfh^{\ell+1,e} \in \mathbb{F}_{2^{3c\log m}}^{N^{\ell+1}}$. Then, there are at most $2\left( t+t_e \right)$ and $2t_e $ erased and erroneous symbols (over the field $ \mathbb{F}_{2^{3c\log m}}$) in $\bfh^{\ell+1,e}$, respectively.
Let $\bfr^{\ell+1}\in \mathbb{F}_{2^{3c\log m}}^{2t+6t_e}$ be the redundancy corresponding to level $\ell+1$ added to $\bfr^B$ in Step 4. Then, applying the Reed Solomon decoder on $\bfh^{\ell+1,e}| \bfr^{\ell+1}$ recovers $\bfh^{\ell+1}$. We proceed in a similar way to retrieve the hash values of the next levels.

Claims~\ref{claim:Erasures} and~\ref{claim:Erasures} follow from the papers~\cite{DocExchange_Many,DocExchange_Hauepler}, and~\cite{IrmakDocExch}.

\begin{claim} \label{claim:Erasures}
The longest matching satisfies $\Tilde{k} \geq N - t-t_e$.  
\end{claim}

\begin{claim} \label{claim:Errors}
There are at most $t_e$ erroneous blocks in $\bfz^{\ell,e}$.
\end{claim}

\end{proof}

In the papers~\cite{DocExchange_Many} and~\cite{DocExchange_Hauepler}, to correct insertions and deletions, a longest matching between a codeword and a single noisy word is sought. To accommodate breakings, we find matching here with all possible concatenations. Hence, the decoding run-time complexity is $\Theta \left(t! n^4 \right)$, as there are $O(n)$ levels, and the Reed-Solomon decoder, and finding a matching (by Lemma 3.2 in the paper~\cite{DocExchange_Many}), require $\Theta \left( n^3\right)$ run-time complexity each.

\section{Proof of Theorem~\ref{theorem:ExplicitCode}} \label{sec:Explicit}

Construction~\ref{Cons:ExplicitBreaksErrorsCode} modifies Construction~\ref{Cons:BreaksErrorsCode} to propose a polynomial run-time encoder that generates a $t$-breaks $t_e$-edit-errors resilient code ${\cC}^{b,e,enc}_{t,t_e}$. To that end, the following lemma, which constructs a mutually uncorrelated code (as defined in the following) will be used.
A mutually uncorrelated code is a set of codewords such that the prefix of one
codeword does not coincide with the suffix of any (potentially
identical) other codeword~\cite{GilbertMU, LevenshteinMU1, LevenshteinMU2, LevenshteinMU3, Maya}. Lemma~\ref{lem:0I0MUCode} provides a mutually uncorrelated code similar to the one used in the work~\cite{Raviv2024_Practical}.

\begin{lemma} \label{lem:0I0MUCode}
For $\ell_1,\ell_2 \in \mathbb{N}$, where $\ell_1 > \ell_2$, let
$$\cC_{MU} \left( \ell_1, \ell_2\right) \triangleq \left\{ 0^{\ell_1} | 1 | \bfw | 1 : \bfw \in \left\{0,1 \right\}^{\ell_2} \right\},
$$
where $0^{\ell_1}$ denotes a length-$\ell_1$ run of $0$'s.
Then,
$\cC_{MU}\left(\ell_1, \ell_2 \right)$ is a mutually uncorrelated code.
\end{lemma}

\begin{cons} \label{Cons:ExplicitBreaksErrorsCode}
    Let $\bfz \in \{0,1\}^m$. The following steps generate a codeword $\bfc \hspace{-0.4ex} \in \hspace{-0.4ex} \cC^{b,e,enc}_{t,t_e}$ in a polynomial run-time complexity.

    \noindent \textbf{Step 1:} Let $\ell_2 = \left\lceil \log \left( 36 \left(t+t_e \right) \log \log m \right) \right\rceil $, and $\ell_1$ be the smallest integer that satisfies
    the inequality $ \ell_1 \geq \max \left\{ 2\log m+2, \ell_2\right\} $
    and also ensures that $ \ell_1+\ell_2+2$ is a multiple of $\log m$ greater than $5\log m$. Then, let $c$ be such that $\ell_1 \hspace{-0.2ex}+\hspace{-0.2ex} \ell_2 \hspace{-0.2ex}+ \hspace{-0.2ex}2 \hspace{-0.2ex} = \hspace{-0.2ex} c\log m$ and $M_{red} \hspace{-0.2ex}=\hspace{-0.2ex} \cC_{MU}\left( \ell_1,\ell_2 \right)$ (see Lemma~\ref{lem:0I0MUCode}).

    \noindent\textbf{Step 3:} Apply the repeat-free algorithm devised by Elishco et al.~\cite{RFCodes} to obtain a length-$(m+2)$ $\left(2\log n+2 \right)$-repeat-free word $\bfz'$.

    \noindent\textbf{Step 4:} Level-$1$ blocks are obtained by dividing $\bfz'$ evenly to $t_e+t+1$ blocks, where all blocks are necessarily of length $\left\lfloor \frac{m}{t+t_e+1}\right\rfloor$, except for the last one which might be shorter. keep dividing the blocks as in Subsection~\ref{subsec:Randominzed} until all blocks are of length at most $3c\log m$. The hash value at all levels now is the length-$\left(3c\log m \right)$ prefix of each block.

    \noindent\textbf{Step 5:} Do not construct redundancy $\bfr^A$ defined in Subsection~\ref{subsec:Literature}. Instead, assign $\bfr^B = \bfh^1_1 | \cdots | \bfh^1_{N^{1}}$. Then append to $\bfr^B$ the redundancy on the hash values of the upcoming levels as in Step $4$ in Construction~\ref{Cons:BreaksErrorsCode}, yet now produce $20t + 20t_e$ redundant symbols over the field $\mathbb{F}_{2^{3 c\log m}}$ for each level (instead of $2t+6t_e$ symbols in Construction~\ref{Cons:BreaksErrorsCode}). At the end, add redundancy $\bfr'$ as in Construction~\ref{Cons:BreaksErrorsCode}.

    \noindent\textbf{Step 6:} 
    To obtain $\bfr$, insert markers from $M_{red}$ in a predetermined order in $\bfr^B$ as in Construction~\ref{Cons:BreaksErrorsCode}.
\end{cons}

\begin{proof} [Sketch proof of Theorem~\ref{theorem:ExplicitCode}] First, one may prove that the added redundancy's length is of order $\Theta \left(\left( t+t_e \right) \log m \log \frac{m}{t+t_e} \right)$.
Next let $\bfz\in \{0,1\}^m$, and let $\bfc = \bfz' | \bfr \in \cC^{b,e,enc}_{t,t_e}$ be the codeword obtained from Construction~\ref{Cons:ExplicitBreaksErrorsCode}, where $\bfz'$ is specifically received at Step $3$. Once $\bfz'$ is found, $\bfz$ can be recovered using the decoder in the paper~\cite{RFCodes}. Thus, we recover $\bfz'$ as follows.

Denote the received noisy fragments by $\bff_1, \ldots, \bff_{t+1}$. 
First recover $\bfr$ exactly as in Step 1 in the proof of Theorem~\ref{theorem:RandomCode}. This immediately gives $\bfh^1_1, \ldots, \bfh^{1}_{N^{\ell}}$ and hence one can continue to iteratively recover the hash values of all upcoming levels (of $\bfz'$) as in Step 3 in the proof of Theorem~\ref{theorem:RandomCode}. The correctness of the decoding is guaranteed for 
$\bfz'$ does not contain $M_{red}$ markers and $\bfc$ is $(3c\log m)$-repeat-free. The run-time of this reconstruction process is $\Theta \left( t! n^4 \right)$ as in Subsection~\ref{subsec:Randominzed}.

Next assume only substitutions occurred in the fragments. Now given $\bfh^{\ell} = \left( \bfh^{\ell}_1, \cdots , \bfh^{\ell}_{N^{\ell}}\right)$ to find $\bfh^{\ell+1} = \left( \bfh^{\ell+1}_1, \cdots , \bfh^{\ell+1}_{N^{\ell+1}}\right)$ we will not search for a matching with all concatenations. Instead we recover $\bfh^{\ell+1}$ as follows. 
For fragment $\bff_i$ and index $j$, as illustrated in Fig.~\ref{fig:Affixing}, define $N \left( \bff_i,j \right)$ to be the number of hash values that would match if we affix $\bff_i$ at index $j$. Moreover, let $J\left( \bff_i \right)$ be the set of indices $j$ such that if we affix $\bff_i$ at index $j$, then $\bff_i$ agrees on at least half of the traversed hash values.
Lastly, define 

    $$J_{max} \left( \bff_i\right) = \begin{cases}
        -1, & J \left(\bff_i\right) = \emptyset \\
        \arg\max_{j\in J \left( \bff_i\right)} N \left( \bff_i, j\right), & \text{otherwise}
    \end{cases} .$$
    Intuitively, $J_{max} \left( \bff_i\right)$ is a guess of the right position of $\bff_{i}$.
    Next, arrange the fragments in order $\bff_{i_1} \hspace{-0.3ex}, \hspace{-0.3ex} \bff_{i_2} \hspace{-0.3ex},  \hspace{-0.3ex} \ldots \hspace{-0.3ex}, \hspace{-0.3ex}  \bff_{i_{t+1}}$ so that $J_{max} \left(\bff_{i_1} \right) \leq J_{max}\left( \bff_{i_2} \right) \leq \cdots \leq J_{max} \left( \bff_{i_{t+1}}\right)$ to iterate over them in this order, while ignoring all fragments for which $J_{max}(\cdot)$ equals $-1$. Iterate over the fragments, and for fragment $\bff_{i_h}$, affix it at index $J_{\max} \left( \bff_{i_h}\right)$ if there does not exist any other fragment $\bff_{i_{h'}}$ that follows $\bff_h$ in the order above and satisfies 
    $$ J_{max} \left( \bff_{i_{h}} \right) \leq  J_{max} \left( \bff_{i_{h'}} \right) \leq J_{max} \left( \bff_{i_{h}} \right) + \left| \bff_{i_{h}} \right| - 1$$ 
        (i.e., affixing $\bff_{i_{h'}}$ at $J_{\max} \left( \bff_{i_{h'}}\right)$ would overlap with affixing $\bff_{i_h}$ at $J_{\max} \left( \bff_{i_h}\right)$) and  
        $N\left( \bff_{i_{h'}}, J_{max}\left( \bff_{i_{h'}} \right) \right) > N\left( \bff_{i_{h}}, J_{max}\left( \bff_{i_{h}} \right) \right)$.
    Otherwise, if there exists such $\bff_{i_{h'}}$, do not affix $\bff_{i_h}$ and continue iterating over the fragments.

At the end of these iterations some fragments are affixed to some positions, and hence provide estimation of the content of $\bfc$ at some positions. From this estimation of $\bfc$ we derive an estimation $\bfh^{\ell+1,e}$ of $\bfh^{\ell+1}$ (as in the proof of Theorem~\ref{theorem:RandomCode}). One can show that there are at most $4t+4t_e$ and $12t+12t_2$ errors and erasures in $\bfh^{\ell+1,e}$, respectively. Thus, we recover $\bfh^{\ell+1}$ using the Reed-Solomon decoder.

\begin{figure}
\centering
\includegraphics[scale=0.5]{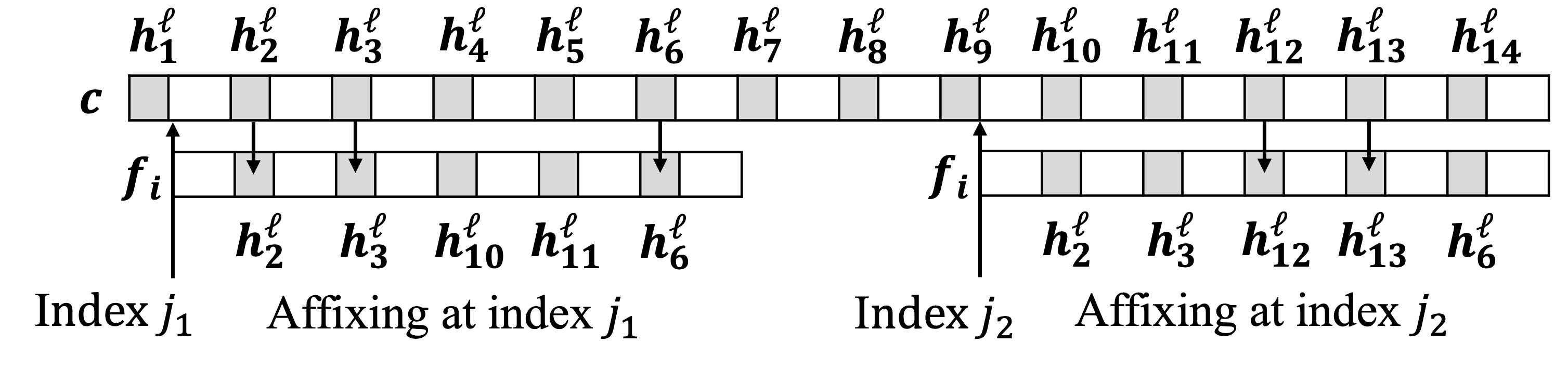}    

\caption{$ N\left(\bff_i,j_1 \right)=3, N \left( \bff_i, j_2\right) = 2  $. Moreover, $j_1 \in J \left(\bff_i\right)$ as half of the $5$ traversed hash values match, while $j_2 \notin J \left(\bff_i\right)$.}
\label{fig:Affixing}
\end{figure}

\end{proof}

\section{List Decoding of the Torn Paper Channel} \label{Sec:ListDecod}

This section focuses on the error-free torn paper channel, hence given a word $\bfw\in \{0,1\}^*$, we define $B^b_t\left(\bfw\right)$ to be the set of all outputs that may be received when transmitting $\bfw$ over an error-free $t$-breaks channel. Thus, $B^b_t(\bfw)$ is a set of sets of fragments. Let $\cC$ be a length-$n$ $t$-breaks resilient code. Then, for every distinct codewords $\bfc, \bfc' \in \cC$, it is guaranteed that $B^b_t \left(\bfc \right) \cap B^b_t\left( \bfc'\right) = \emptyset$, yet we might have
$ B^b_{t'} \left(\bfc \right) \cap B^b_{t'}\left( \bfc'\right) \neq \emptyset $ for $t' > t$. Thus, given the output $S $ of transmitting some codeword of $\cC$ over a $t'$-breaks channel, it might be impossible to determine the specific transmitted codeword. In this case a list of all potential codewords 
$ \left\{ \bfc \in \cC : S \in B^b_{t'} \left(\bfc\right) \right\} $
may be generated. This section investigates the maximum possible size of such a list, defined by
$$ L\left(t,t',n \right) \triangleq \max_{\bfc, \bfc' \in \{0,1\}^n : B^b_t \left( \bfc \right) \cap B^b_{t}\left( \bfc' \right) = \emptyset }  \left| B^b_{t'} \left( \bfc \right) \cap B^b_{t'} \left( \bfc' \right) \right|. $$
Notice that $L\left(t,t',n\right) \leq \left(t'+1 \right)!$, for the codewords are obtained by some concatenations of the $t'+1$ fragments. Theorem~\ref{theorem:ListDecUpBound1} improves this observation.

The key idea behind Theorem~\ref{theorem:ListDecUpBound1} is that a potential codeword $\bfx$ may be obtained by some concatenation of the fragments, such as $\bff_{i_1}| \bff_{i_2}| \cdots | \bff_{i_{t+1}}$. Thus, assume we have a decoder $\cD_{\cC}$ of the code $\cC$, which gets a set of $t+1$ fragments and finds whether there exists a codeword corresponding to the $t+1$ fragments. If there exists a suitable codeword, $\cD$ outputs it as well. Then, we can recover $\bfc$ by executing $\cD$ on the set of fragments
$ \left\{ \bff_{i_1}| \cdots | \bff_{i_{t'-t+1}}  \right\} \cup \left\{ \bff_{i_j}: t'-t+2\leq j \leq t'+1 \right\}. $ Similarly, $\bfc$ may be obtained from concatenating the second or the third length-$\left( t'-t+1\right)$ sequence of fragments instead. This intuition may be developed to derive Theorem~\ref{theorem:ListDecUpBound1}. 

\begin{theorem} \label{theorem:ListDecUpBound1}
For every $t,t',n\in \mathbb{N}$ where $t<t'$, it holds that
$$L \left(t,t',n\right) \leq \binom{t'+1}{t'-t+1} \frac{\left( t'-t+1 \right)!}{t+1} .$$
\end{theorem}

By Theorem~\ref{theorem:ListDecUpBound1}, if $t'=t+c$ for some constant $c$, i.e., only few more breakings happened compared to the correcting capability of $\cC$, then $L\left(t,t',n \right) = O \left(t^{c-1} \right)$. 
Lastly, Theorem~\ref{theorem:ListDecLowBound} establishes a lower bound on $L\left(t,t',n \right)$.

\begin{theorem} \label{theorem:ListDecLowBound}
    For $t,t',n$ such that $t' \geq \max \left\{t,6 \right\}$ and 
    $$n\geq \left(t'+1\right)\left(3+ 2\left\lceil \log \left( 2\left(t'+1 \right) \right)\right\rceil \right),$$ 
    it holds that $ L \left(t,t',n \right) \geq t'+1$.
\end{theorem}

Theorem~\ref{theorem:ListDecLowBound} is established using Lemma~\ref{lem:0I0MUCode} and Hamiltonian decomposition of complete directed graphs~\cite{EtzionHamiltonian, BFHamiltonian, TillsonHamiltonian}. The proof is omitted due to the lack of space.

\section*{Acknowledgements}

 Funded by the European Union (DiDAX, 101115134). Views and opinions expressed are however those of the
authors only and do not necessarily reflect those of the European Union
or the European Research Council Executive Agency. Neither the European
Union nor the granting authority can be held responsible for them.

\end{document}